\documentclass{article}
\usepackage{bm}
\begin{document}

\title{NONZERO $\theta_{13}$ AND NEUTRINO MASSES FROM MODIFIED NEUTRINO MIXING MATRIX}

\maketitle

\begin{center}
\textbf ASAN DAMANIK\\
\it Faculty of Science and Technology,\\Sanata Dharma University,\\Kampus III USD Paingan Maguwoharjo Sleman, Yogyakarta, Indonesia\\d.asan@lycos.com
\end{center}
\begin{abstract}
The nonzero and relatively large $\theta_{13}$ have been reported by Daya Bay, T2K, MINOS, and Double Chooz Collaborations. In order to accommodate the nonzero $\theta_{13}$, we modified the tribimaximal (TB), bimaxima (BM), and democratic (DC) neutrino mixing matrices.  From three modified neutrino mixing matrices, two of them (the modified BM and DC mixing matrices) can give nonzero $\theta_{13}$ which is compatible with the result of the Daya Bay and T2K experiments. The modified TB neutrino mixing matrix predicts the value of $\theta_{13}$ greater than the upper bound value of the latest experimental results.  By using the modified neutrino mixing matrices and impose an additional assumption that neutrino mass matrices have two zeros texture, we then obtain the neutrino mass in normal hierarchy when $(M_{\nu})_{22}=(M_{\nu})_{33}=0$ for the neutrino mass matrix from the modified TB neutrino mixing matrix and $(M_{\nu})_{11}=(M_{\nu})_{13}=0$ for the neutrino mass matrix from the modified DC neutrino mixing matrix. For these two patterns of neutrino mass matrices,  either the atmospheric mass squared difference or the solar mass squared difference can be obtained, but not both of them simultaneously.  From four patterns of two zeros texture to be considered on the obtained neutrino mass matrix from the modified BM neutrino mixing matrix, none of them can predict correctly neutrino mass spectrum (normal or inverted hierarchy).

\begin{flushleft}
\end{flushleft}
\end{abstract}

\section{Introduction}

Recently, there is a convincing evidence that neutrinos have a non-zero mass.  This evidence is based on the experimental facts that both solar and atmospheric neutrinos undergo oscillations.\cite{Fukuda98}-\cite{Fukugita03}  Since neutrinos are massive, there will be flavor mixing in the charged current interactions of the leptons and a leptonic mixing matrix will appear analogous to the mixing matrix in quarks sector.  The mixing matrix in neutrino sector links the mass eigenstates of neutrino $(\nu_{1}, \nu_{2}, \nu_{3})$ to the flavor eigenstates of neutrino $(\nu_{e}, \nu_{\mu}, \nu_{\tau})$ as follow:
\begin{equation}
\bordermatrix{& \cr
&\nu_{e}\cr
&\nu_{\mu}\cr
&\nu_{\tau}\cr}=V\bordermatrix{& \cr
&\nu_{1}\cr
&\nu_{2}\cr
&\nu_{3}\cr}
 \label{V}
\end{equation}
where $V$ is the $3\times 3$ neutrino mixing matrix.

The neutrino mixing matrix $V$, which is also known as PMNS matrix\cite{Pontecorvo58, Maki}, contains three mixing angles and three CP violating phases (one Dirac type and two Majorana type).  In the standard parametrization the neutrino mixing matrix $V$ is given by:
\begin{equation}
V=\bordermatrix{& & &\cr
&c_{12}c_{13} &s_{12}c_{13} &z^{*}\cr
&-s_{12}c_{23}-c_{12}s_{23}z &c_{12}c_{23}-s_{12}s_{23}z &s_{23}c_{13}\cr
&s_{12}s_{23}-c_{12}c_{23}z &-c_{12}s_{23}-s_{12}c_{23}z &c_{23}c_{13}\cr},
 \label{V1}
\end{equation}
where $c_{ij}$ and $s_{ij}$ stand for $\cos\theta_{ij}$ and $\sin\theta_{ij}$ respectively, and $z=s_{13}e^{i\varphi}$.

From the theoretical point of view, there are three well-known patterns of neutrino mixing matrix $V$: tribimaximal mixing pattern (TB)\cite{Harrison}-\cite{He}, bimaximal mixing pattern (BM)\cite{Vissani}-\cite{Li}, and democratic mixing pattern(DC).\cite{Fritzsch96}-\cite{Fritzschb}  Explicitly, the form of the neutrino mixing matrices read:
\begin{equation}
V_{\rm{TB}}=\bordermatrix{& & &\cr
&\sqrt{\frac{2}{3}} &\sqrt{\frac{1}{3}} &0\cr
&-\sqrt{\frac{1}{6}} &\sqrt{\frac{1}{3}} &\sqrt{\frac{1}{2}}\cr
&-\sqrt{\frac{1}{6}} &\sqrt{\frac{1}{3}} &-\sqrt{\frac{1}{2}}\cr},~V_{\rm{BM}}=\bordermatrix{& & &\cr
&\sqrt{\frac{1}{2}} &\sqrt{\frac{1}{2}} &0\cr
&-\frac{1}{2} &\frac{1}{2} &\sqrt{\frac{1}{2}}\cr
&\frac{1}{2} &-\frac{1}{2} &\sqrt{\frac{1}{2}}\cr},\nonumber
\end{equation}
\begin{equation}
V_{\rm{DC}}=\bordermatrix{& & &\cr
&\sqrt{\frac{1}{2}} &\sqrt{\frac{1}{2}} &0\cr
&\sqrt{\frac{1}{6}} &-\sqrt{\frac{1}{6}} &-\sqrt{\frac{2}{3}}\cr
&-\sqrt{\frac{1}{3}} &\sqrt{\frac{1}{3}} &-\sqrt{\frac{1}{3}}\cr},
 \label{tb}
\end{equation}
which lead to $\theta_{13}=0$.  However, the latest result from long baseline neutrino oscillation experiment T2K indicates that $\theta_{13}$ is relatively large.  For a vanishing Dirac CP-violating phase, the T2K collaboration reported that the values of $\theta_{13}$ for neutrino mass in normal hierarchy (NH) are\cite{T2K}:
\begin{equation}
5.0^{o}\leq\theta_{13}\leq 16.0^{o},
\end{equation}
and 
\begin{equation}
5.8^{o}\leq\theta_{13}\leq 17.8^{o},
\end{equation}
for neutrino mass in inverted hierarchy (IH), and the current combined world data\cite{Gonzales-Carcia}-\cite{Fogli}:
\begin{equation}
\Delta m_{21}^{2}=7.59\pm0.20 (_{-0.69}^{+0.61}) \times 10^{-5}~\rm{eV^{2}},\label{21}
\end{equation}
\begin{equation}
\Delta m_{32}^{2}=2.46\pm0.12(\pm0.37) \times 10^{-3}~\rm{eV^{2}},~\rm(for~ NH)\label{32}
\end{equation}
\begin{equation}
\Delta m_{32}^{2}=-2.36\pm0.11(\pm0.37) \times 10^{-3}~\rm{eV^{2}},~\rm(for~ IH)\label{321}
\end{equation}
\begin{equation}
\theta_{12}=34.5\pm1.0 (_{-2.8}^{3.2})^{o},~~\theta_{23}=42.8_{-2.9}^{+4.5}(_{-7.3}^{+10.7})^{o},~~\theta_{13}=5.1_{-3.3}^{+3.0}(\leq 12.0)^{o},
 \label{GD}
\end{equation}
at $1\sigma~(3\sigma)$ level.  The latest experimental result on $\theta_{13}$ is reported by Daya Bay Collaboration which gives\cite{Daya}:
\begin{equation}
\sin^{2}2\theta_{13}=0.092\pm 0.016 (\rm{stat}.)\pm 0.005 (\rm{syst.}).
\end{equation}

In order to accommodate nonzero $\theta_{13}$ value, several models and modification of neutrino mixing matrix have been proposed by many authors. By analyzing two-zero texture of neutrino mass matrix ($M_{\mu\mu}=M_{\tau\tau}=0$), very large $\theta_{13}$ can be produced if atmospheric neutrino oscillations are not too nearly maximal.\cite{Frampton02}-\cite{Ludl}    Nonzero $\theta_{13}$ in the context of $A_{4}$ model was discussed in Refs.~\cite{Ma10}\--\cite{King}, and by using $S_{4}$ flavor symmetry with leaving maximal $\theta_{23}$ and trimaximal $\theta_{12}$ was discussed in Refs.~\cite{Morisi11}--\cite{Chu}.  Relatively large $\theta_{13}$ can also be obtained by applying permutation symmetry $S_{3}$ to both charged lepton and neutrino mass matrices in which the flavor symmetry is explicitly broken down with different symmetry breaking.\cite{Zhou11}  Minimal modification to the neutrino mixing matrix (tribimaximal, bimaximal, and democratic) can be found in Refs.~\cite{Xing11}--\cite{Wchao}, nonzero $\theta_{13}$ and CP-violation in inverse neutrino mass matrix with two texture zero is discussed in Refs.~\cite{Verma11}--\cite{Rodejohann11}.    By using the criterion that the mixing matrix can be parameterized by three rotation angles which are simple fraction of $\pi$, there are twenty successful mixing patterns to be consistent with the latest neutrino oscillation data.\cite{Rodejohann11}  The non-zero $\theta_{13}$ can also be derived from a Super-symmetric $B-L$ gauge model with $T_{7}$ lepton flavor symmetry, SO(10) with type II seesaw, finite quantum correction in quasi-degenerate neutrino mass spectrum, and introducing a small correction term $\delta M$  in the neutrino sector (see Refs.~ \cite{Cao11}--\cite{Araki}).

 Neutrino mixing matrix can be used to obtain neutrino mass matrix.  One of the interesting patterns of neutrino mass matrix that have been extensively studied in literature is the texture zero.  Neutrino mass matrix with texture zero is a consequence of the underlying symmetry in a given model and is phenomenologically useful in the sense that they guarantee the calculability of $M_{\nu}$ from which both the neutrino mass spectrum and the flavor mixing pattern can more or less be predicted \cite{Fritzsch}.  In view of the latest T2K neutrino oscillation data which hint a relatively large $\theta_{13}$ in relation to texture zero of neutrino mass, Kumar\cite{Kumar} discussed the implications of a class of neutrino mass matrices with texture zero and allows the deviations from maximal mixing, Deepthi {\it et al.}\cite{Deepthi11} analyzed one texture zero of neutrino mass matrix, and  Fritzsch {\it et al.}\cite{Fritzsch} performed a systematic study of the neutrino mass matrix with two independent texture zero.

In this paper, we use the modified neutrino mixing matrix (TB, BM, and DC) in order to obtain nonzero $\theta_{13}$ similar to the Deepthi {\it et al.} paper\cite{Deepthi11}, but with different zero texture.  We use the modified neutrino mixing matrices to obtain the neutrino mass matrices which have two zeros texture.  The neutrino mass and its hierarchies from the obtained neutrino mass matrices are studied systematically and discuss its phenomenological consequences.  This paper is organized as follow: in section 2, the modified neutrino mixing matrices (TB, BM, and DC) to be reviewed and in section 3 the neutrino mass matrices from modified neutrino mixing matrix with two zeros texture to be constructed and discuss its phenomenological consequences.  Finally, section 4 is devoted to conclusions. 

\section{Modified Neutrino Mixing Matrices}

As we have stated explicitly in section 1, in this section we modify the tribimaximal, bimaximal, and democratic neutrino mixing matrices patterns in Eq. (\ref{tb}).  Modification of neutrino mixing matrix, by introducing  perturbation matrices into neutrino mixing matrices in Eq. (\ref{tb}), is the easiest way to obtain the nonzero $\theta_{13}$.  The value of $\theta_{13}$ can be obtained in some parameters that can be fitted from experimental results.  In this paper, the modified neutrino mixing matrices to be considered are given by:
\begin{equation}
V_{{\rm TB}}^{'}=V_{{\rm TB}}V_{23}V_{12},\label{Modi1}
\end{equation}
\begin{equation}
V_{{\rm BM}}^{'}=V_{{\rm BM}}V_{23}V_{12},\label{Modi2}
\end{equation}
\begin{equation}
V_{{\rm DC}}^{'}=V_{{\rm DC}}V_{23}V_{12},
 \label{Modified}
\end{equation}
where $V_{12}$ and $V_{23}$ are the perturbation matrices to the neutrino mixing matrices.  We take the form of the perturbation matrices as follow:
\begin{equation}
V_{12}=\bordermatrix{& & &\cr
&c_{x} &s_{x} &0\cr
&-s_{x} &c_{x} &0\cr
&0 &0 &1\cr},~V_{23}=\bordermatrix{& & &\cr
&1 &0 &0\cr
&0 &c_{y} &s_{y}\cr
&0 &-s_{y} &c_{y}\cr}.
 \label{xy}
\end{equation}
where $c_{x}$ is the $\cos{x}$, $s_{x}$ is the $\sin{x}$, $c_{y}$ is the $\cos{y}$, and $s_{y}$ is the $\sin{y}$.

By inserting Eqs. (\ref{tb}) and (\ref{xy}) into Eqs. (\ref{Modi1})-(\ref{Modified}), we then have the modified neutrino mixing matrices as follow:
\begin{equation}
V_{{\rm TB}}^{'}=\bordermatrix{& & &\cr
&\frac{\sqrt{3}}{3}(\sqrt{2}c_{x}-c_{y}s_{x}) &\frac{\sqrt{3}}{3}(\sqrt{2}s_{x}+c_{y}c_{x}) &\frac{\sqrt{3}}{3}s_{y}\cr
&-\frac{\sqrt{3}}{3}(\frac{\sqrt{2}}{2}c_{x}+c_{y}s_{x})+\frac{\sqrt{2}}{2}s_{y}s_{x} &-\frac{\sqrt{3}}{3}(\frac{\sqrt{2}}{2}s_{x}-c_{y}c_{x})-\frac{\sqrt{2}}{2}s_{y}c_{x} &\frac{\sqrt{3}}{3}s_{y}+\frac{\sqrt{2}}{2}c_{y}\cr
&-\frac{\sqrt{3}}{3}(\frac{\sqrt{2}}{2}c_{x}+c_{y}s_{x})-\frac{\sqrt{2}}{2}s_{y}s_{x} &-\frac{\sqrt{3}}{3}(\frac{\sqrt{2}}{2}s_{x}-c_{y}c_{x})+\frac{\sqrt{2}}{2}s_{y}c_{x} &\frac{\sqrt{3}}{3}s_{y}-\frac{\sqrt{2}}{2}c_{y}\cr},\label{Mo1}
\end{equation}
\begin{equation}
V_{{\rm BM}}^{'}=\bordermatrix{& & &\cr
&\frac{\sqrt{2}}{2}(c_{x}-c_{y}s_{x}) &\frac{\sqrt{2}}{2}(s_{x}+c_{y}c_{x}) &\frac{\sqrt{2}}{2}s_{y}\cr
&-\frac{1}{2}(c_{x}+c_{y}s_{x}-\sqrt{2}s_{y}s_{x}) &-\frac{1}{2}(s_{x}-c_{y}c_{x}+\sqrt{2}s_{y}c_{x}) &\frac{1}{2}(s_{y}+\sqrt{2}c_{y})\cr
&\frac{1}{2}(c_{x}+c_{y}s_{x}+\sqrt{2}s_{y}s_{x}) &\frac{1}{2}(s_{x}-c_{y}c_{x}-\sqrt{2}s_{y}c_{x}) &-\frac{1}{2}(s_{y}-\sqrt{2}c_{y})\cr},\label{Mo2}
\end{equation}
\begin{equation}
V_{{\rm DC}}^{'}=\bordermatrix{& & &\cr
&\frac{\sqrt{2}}{2}(c_{x}-c_{y}s_{x}) &\frac{\sqrt{2}}{2}(s_{x}+c_{y}c_{x}) &\frac{\sqrt{2}}{2}s_{y}\cr
&\frac{\sqrt{6}}{6}(c_{x}+c_{y}s_{x}-2s_{y}s_{x}) &\frac{\sqrt{6}}{6}(s_{x}-c_{y}c_{x}+2s_{y}c_{x}) &-\frac{\sqrt{6}}{6}(s_{y}+2c_{y})\cr
&-\frac{\sqrt{3}}{3}(c_{x}+c_{y}s_{x}+s_{y}s_{x}) &-\frac{\sqrt{3}}{3}(s_{x}-c_{y}c_{x}-s_{y}c_{x}) &\frac{\sqrt{3}}{3}(s_{y}-c_{y})\cr}.
 \label{Mo}
\end{equation}
By comparing Eqs. (\ref{Mo1}), (\ref{Mo2}), and (\ref{Mo}) with the neutrino mixing in standard parameterization form as shown in Eq. (\ref{V1}) with $\varphi=0$, then we obtain:
\begin{equation}
\tan\theta_{12}=\left|\frac{\sqrt{2}s_{x}+c_{y}c_{x}}{\sqrt{2}c_{x}-c_{y}s_{x}}\right|,~~
\tan\theta_{23}=\left|\frac{\frac{\sqrt{3}}{3}s_{y}+\frac{\sqrt{2}}{2}c_{y}}{\frac{\sqrt{3}}{3}s_{y}-\frac{\sqrt{2}}{2}c_{y}}\right|,~~
\sin\theta_{13}=\left|\frac{\sqrt{3}}{3}s_{y}\right|,
 \label{1}
\end{equation}
for modified tribimaximal mixing, and
\begin{equation}
\tan\theta_{12}=\left|\frac{s_{x}+c_{y}c_{x}}{c_{x}-c_{y}s_{x}}\right|,~~
\tan\theta_{23}=\left|-\frac{s_{y}+\sqrt{2}c_{y}}{s_{y}-\sqrt{2}c_{y}}\right|,~~
\sin\theta_{13}=\left|\frac{\sqrt{2}}{2}s_{y}\right|,
 \label{2}
\end{equation}
for modified bimaximal mixing, and
\begin{equation}
\tan\theta_{12}=\left|\frac{s_{x}+c_{y}c_{x}}{c_{x}-c_{y}s_{x}}\right|,~~
\tan\theta_{23}=\left|-\frac{\sqrt{2}}{2}\left(\frac{s_{y}+2c_{y}}{s_{y}-c_{y}}\right)\right|,~~
\sin\theta_{13}=\left|\frac{\sqrt{2}}{2}s_{y}\right|,
 \label{3}
\end{equation}
for modified democratic mixing. It is apparent that for $y\rightarrow 0$, the value of $\tan\theta_{23}\rightarrow 1$ for both modified TB and BM, meanwhile for modified DC when $y\rightarrow 0$ the value of the $\tan\theta_{23}\rightarrow \sqrt{2}$.   From  Eqs. (\ref{1}), (\ref{2}), and (\ref{3}), one can see that it is possible to determine the value of $x$ and $y$ and therefore the value of $\theta_{13}$ by using the experimental values of $\theta_{12}$ and $\theta_{23}$.

By inserting the experimental values of $\theta_{12}$ and $\theta_{23}$ in Eq. (\ref{GD}) into Eqs. (\ref{1}), (\ref{2}), and (\ref{3}), we obtain:
\begin{equation}
x\approx 32.21^{o},~~y\approx -88.22^{o}, ~~\rm{for~modified~TB},\label{xtb}
\end{equation}
\begin{equation}
x\approx 45.01^{o},~~y\approx -3.14^{o}, ~~\rm{for~modified~BM},\label{xbm}
\end{equation}
\begin{equation}
x\approx -9.22^{o},~~y\approx -16.68^{o}, ~~\rm{for~modified~DC}\label{xdc},
\end{equation}
which it imply that:
\begin{equation}
\theta_{13}\approx 35.06^{o},~~\rm{for~modified~TB},\label{TB}
\end{equation}
\begin{equation}
\theta_{13}\approx 2.22^{o},~~\rm{for~modified~BM},\label{BM}
\end{equation}
\begin{equation}
\theta_{13}\approx 11.71^{o},~~\rm{for~modified~DC}.\label{DC}
\end{equation}

The values of $x$ and $y$ for both modified TB and BM are in the range of the given values of Ref.~\cite{Deepthi11}, whereas the values of $x$ and $y$ for the modified DC in this paper did not already reported in Ref.~\cite{Deepthi11}.  The modified neutrino mixing matrices, within the scheme of Eqs. (\ref{Modi1})-(\ref{Modified}), can produce non-zero $\theta_{13}$, but only the modified DC neutrino mixing matrices can predict the values of $\theta_{13}$ that are compatible with the T2K result.  The relatively large $\theta_{13}$ can be obtained from bimaximal neutrino mixing matrix (BM) with specific discrete models, see for example Refs.~\cite{Altarelli}--\cite{Tooropa}. 

\section{Neutrino Mass Matrix and Neutrino Mass Spectrum}

In this section, we analyze the predictions of all the modified neutrino mixing matrices on neutrino mass and its neutrino mass spectrum because all of them can predict the nonzero $\theta_{13}$.  The neutrino mass matrix to be constructed by using the modified neutrino mixing matrix that have already been reviewed in section 2 and impose an additional assumption: that obtained neutrino mass matrix has two zeros texture.  Four patterns of two zeros textures to be considered in the obtained neutrino mass matrix are the zero textures:
\begin{eqnarray}
(M_{\nu})_{22}=(M_{\nu})_{33}=0,\label{2233}\\
(M_{\nu})_{11}=(M_{\nu})_{13}=0,\label{1113}\\
(M_{\nu})_{12}=(M_{\nu})_{13}=0,\label{1213}\\
(M_{\nu})_{12}=(M_{\nu})_{23}=0.\label{1223}
\end{eqnarray}

We construct the neutrino mass matrix in flavor eigenstates basis (where the charged lepton mass matrix is diagonal).  In this basis, the neutrino mass matrix can be diagonalized by a unitary matrix $V$ as follow:
\begin{equation}
M_{\nu}=V M V^{T},
 \label{Mf}
\end{equation}
where the diagonal neutrino mass matrix $M$ is given by:
\begin{equation}
M=\bordermatrix{& & &\cr
&m_{1} &0 &0\cr
&0 &m_{2} &0\cr
&0 &0 &m_{3}\cr}.
 \label{Mb}
\end{equation}
If the unitary matrix $V$ is replaced by $V_{TB}^{'}$, $V_{BM}^{'}$, or $V_{DC}^{'}$, then Eq. (\ref{Mf}) becomes:
\begin{equation}
M_{\nu}=V_{\alpha}^{'} M V_{\alpha}^{'T},
 \label{A}
\end{equation}
 where $\alpha$ is the index for TB, BM, or DC.

\subsection{Neutrino mass matrix from modified TB}
By using Eqs. (\ref{Mo1}), (\ref{Mb}), and (\ref{A}), we have the neutrino mass matrix with modified tribimaximal neutrino mixing matrix as follows:
\begin{equation}
M_{\nu}=\bordermatrix{& & &\cr
&(M_{\nu})_{11} &(M_{\nu})_{12} &(M_{\nu})_{13} \cr
&(M_{\nu})_{21} &(M_{\nu})_{22} &(M_{\nu})_{23} \cr
&(M_{\nu})_{31} &(M_{\nu})_{32} &(M_{\nu})_{33}\cr},
\end{equation}
where
\begin{equation}
(M_{\nu})_{11}=m_{1}\left(\frac{\sqrt{6}c_{x}}{3}-\frac{\sqrt{3}c_{y}s_{x}}{3}\right)^{2}+m_{2}\left(\frac{\sqrt{6}s_{x}}{3}+\frac{\sqrt{3}c_{y}c_{x}}{3}\right)^{2}+m_{3}\frac{s_{y}^{2}}{3},
\end{equation}
\begin{eqnarray}
(M_{\nu})_{12}=m_{1}\left(\frac{\sqrt{6}c_{x}}{3}-\frac{\sqrt{3}c_{y}c_{x}}{3}\right)\left(-\frac{\sqrt{6}c_{x}}{6}-\frac{\sqrt{3}c_{y}s_{x}}{3}+\frac{\sqrt{2}s_{y}s_{x}}{2}\right)\nonumber\\+m_{2}\left(\frac{\sqrt{6}s_{x}}{3}+\frac{\sqrt{3}c_{y}c_{x}}{3}\right)\left(-\frac{\sqrt{6}s_{x}}{6}+\frac{\sqrt{3}c_{y}c_{x}}{3}-\frac{\sqrt{2}s_{y}c_{x}}{2}\right)\nonumber\\+m_{3}\frac{\sqrt{3}}{3}\left(\frac{\sqrt{3}s_{y}^{2}}{3}+\frac{\sqrt{2}s_{y}c_{y}}{2}\right),
\end{eqnarray}
\begin{eqnarray}
(M_{\nu})_{13}=m_{1}\left(\frac{\sqrt{6}c_{x}}{3}-\frac{\sqrt{3}c_{y}s_{x}}{3}\right)\left(-\frac{\sqrt{6}c_{x}}{6}-\frac{\sqrt{3}c_{y}s_{x}}{3}-\frac{\sqrt{2}s_{y}s_{x}}{2}\right)\nonumber\\+m_{2}\left(\frac{\sqrt{6}s_{x}}{3}+\frac{\sqrt{3}c_{y}c_{x}}{3}\right)\left(-\frac{\sqrt{6}s_{x}}{6}+\frac{\sqrt{3}c_{y}c_{x}}{3}+\frac{\sqrt{2}s_{y}c_{x}}{2}\right)\nonumber\\+m_{3}\frac{\sqrt{3}}{3}\left(\frac{\sqrt{3}s_{y}^{2}}{3}-\frac{\sqrt{2}s_{y}c_{y}}{2}\right),
\end{eqnarray}
\begin{eqnarray}
(M_{\nu})_{21}=m_{1}\left(\frac{\sqrt{6}c_{x}}{3}-\frac{\sqrt{3}c_{y}s_{x}}{3}\right)\left(-\frac{\sqrt{6}c_{x}}{6}-\frac{\sqrt{3}c_{y}s_{x}}{3}+\frac{\sqrt{2}s_{y}s_{x}}{2}\right)\nonumber\\+m_{2}\left(\frac{\sqrt{6}s_{x}}{3}+\frac{\sqrt{3}c_{y}c_{x}}{3}\right)\left(-\frac{\sqrt{6}s_{x}}{6}+\frac{\sqrt{3}c_{y}c_{x}}{3}-\frac{\sqrt{2}s_{y}c_{x}}{2}\right)\nonumber\\+m_{3}\frac{\sqrt{3}}{3}\left(\frac{\sqrt{3}s_{y}^{2}}{3}+\frac{\sqrt{2}s_{y}c_{y}}{2}\right),
\end{eqnarray}
\begin{eqnarray}
(M_{\nu})_{22}=m_{1}\left(-\frac{\sqrt{6}c_{x}}{6}-\frac{\sqrt{3}c_{y}s_{x}}{3}+\frac{\sqrt{2}s_{y}s_{x}}{2}\right)^{2}\nonumber\\+m_{2}\left(-\frac{\sqrt{6}s_{x}}{6}+\frac{\sqrt{3}c_{y}c_{x}}{3}-\frac{\sqrt{2}s_{y}c_{x}}{2}\right)^{2}+m_{3}\left(\frac{\sqrt{3}s_{y}}{3}+\frac{\sqrt{2}c_{y}}{2}\right)^{2},
\end{eqnarray}
\begin{eqnarray}
(M_{\nu})_{23}=m_{1}\left(-\frac{\sqrt{6}c_{x}}{6}-\frac{\sqrt{3}c_{y}s_{x}}{3}+\frac{\sqrt{2}s_{y}s_{x}}{2}\right)\left(-\frac{\sqrt{6}c_{x}}{6}-\frac{\sqrt{3}c_{y}s_{x}}{3}-\frac{\sqrt{2}s_{y}s_{x}}{2}\right)\nonumber\\+m_{2}\left(\frac{-\sqrt{6}s_{x}}{6}+\frac{\sqrt{3}c_{y}c_{x}}{3}-\frac{\sqrt{2}s_{y}c_{x}}{2}\right)\left(-\frac{\sqrt{6}s_{x}}{6}+\frac{\sqrt{3}c_{y}c_{x}}{3}+\frac{\sqrt{2}s_{y}c_{x}}{2}\right)\nonumber\\+m_{3}\left(\frac{\sqrt{3}s_{y}}{3}+\frac{\sqrt{2}c_{y}}{2}\right)\left(\frac{\sqrt{3}s_{y}}{3}-\frac{\sqrt{2}c_{y}}{2}\right),
\end{eqnarray}
\begin{eqnarray}
(M_{\nu})_{31}=m_{1}\left(\frac{\sqrt{6}c_{x}}{3}-\frac{\sqrt{3}c_{y}s_{x}}{3}\right)\left(-\frac{\sqrt{6}c_{x}}{6}-\frac{\sqrt{3}c_{y}s_{x}}{3}-\frac{\sqrt{2}s_{y}s_{x}}{2}\right)\nonumber\\+m_{2}\left(\frac{\sqrt{6}s_{x}}{3}+\frac{\sqrt{3}c_{y}c_{x}}{3}\right)\left(-\frac{\sqrt{6}s_{x}}{6}+\frac{\sqrt{3}c_{y}c_{x}}{3}+\frac{\sqrt{2}s_{y}c_{x}}{2}\right)\nonumber\\+m_{3}\frac{\sqrt{3}}{3}\left(\frac{\sqrt{3}s_{y}^{2}}{3}-\frac{\sqrt{2}s_{y}c_{y}}{2}\right),
\end{eqnarray}
\begin{eqnarray}
(M_{\nu})_{32}=m_{1}\left(-\frac{\sqrt{6}c_{x}}{6}-\frac{\sqrt{3}c_{y}s_{x}}{3}+\frac{\sqrt{2}s_{y}s_{x}}{2}\right)\left(-\frac{\sqrt{6}c_{x}}{6}-\frac{\sqrt{3}c_{y}s_{x}}{3}-\frac{\sqrt{2}s_{y}s_{x}}{2}\right)\nonumber\\+m_{2}\left(\frac{-\sqrt{6}s_{x}}{6}+\frac{\sqrt{3}c_{y}c_{x}}{3}-\frac{\sqrt{2}s_{y}c_{x}}{2}\right)\left(-\frac{\sqrt{6}s_{x}}{6}+\frac{\sqrt{3}c_{y}c_{x}}{3}+\frac{\sqrt{2}s_{y}c_{x}}{2}\right)\nonumber\\+m_{3}\left(\frac{\sqrt{3}s_{y}}{3}+\frac{\sqrt{2}c_{y}}{2}\right)\left(\frac{\sqrt{3}s_{y}}{3}-\frac{\sqrt{2}c_{y}}{2}\right),
\end{eqnarray}
\begin{eqnarray}
(M_{\nu})_{33}=m_{1}\left(-\frac{\sqrt{6}c_{x}}{6}-\frac{\sqrt{3}c_{y}s_{x}}{3}-\frac{\sqrt{2}s_{y}s_{x}}{2}\right)^{2}\nonumber\\+m_{2}\left(-\frac{\sqrt{6}s_{x}}{6}+\frac{\sqrt{3}c_{y}c_{x}}{3}+\frac{\sqrt{2}s_{y}c_{x}}{2}\right)^{2}\nonumber\\+m_{3}\left(\frac{\sqrt{3}s_{y}}{3}-\frac{\sqrt{2}c_{y}}{2}\right)^{2},
\end{eqnarray}

If we impose the four patterns of two zeros texture in Eqs. (\ref{2233})-(\ref{1223}) into the neutrino mass matrix that obtained from the modified tribimaximal neutrino mixing matrix and insert the value of $x$ and $y$  in Eq. (\ref{xtb}), then we have:
\begin{eqnarray}
m_{1}=0.737880853~m_{2},~m_{3}=-1.783552908~ m_{2},~{\rm{for}}~(M_{\nu})_{22}=(M_{\nu})_{33}=0, \label{mtb}\\
m_{1}=-1.327010549~m_{2},~m_{3}=1.162476103~ m_{2},~{\rm{for}}~(M_{\nu})_{11}=(M_{\nu})_{13}=0, \label{mtb1}\\
m_{1}=0.9999999995 ~m_{2},~m_{3}= m_{2},~{\rm{for}}~(M_{\nu})_{12}=(M_{\nu})_{13}=0, \label{mtb2}\\
m_{1}=0.9999999993 ~m_{2},~m_{3}=0.9999999999 ~m_{2},~{\rm{for}}~(M_{\nu})_{12}=(M_{\nu})_{23}=0. \label{mtb3}
\end{eqnarray}
From Eqs. (\ref{mtb})-(\ref{mtb3}), it is apparent that only the two zeros texture: $(M_{\nu})_{22}=(M_{\nu})_{33}=0$ can correctly give the neutrino mass spectrum.  From Eq. (\ref{mtb}), we have:
\begin{eqnarray}
\left|\frac{m_{1}}{m_{2}}\right|,~\left|\frac{m_{2}}{m_{3}}\right|<1,
 \label{IH1}
\end{eqnarray}
which predict the normal hierarchy (NH): $\left|m_{1}\right|<\left|m_{2}\right|<\left|m_{3}\right|$.

By using the experimental values of squared mass difference as shown in Eqs. (\ref{21}) into Eq. (\ref{mtb}), we obtain the absolute values of neutrino mass as follow:
\begin{eqnarray}
\left|m_{1}\right|=0.0095246222~\rm{eV},\nonumber\\
\left|m_{2}\right|=0.0129080761~\rm{eV},\nonumber\\
\left|m_{3}\right|=0.0230222366~\rm{eV},\label{MTB}
\end{eqnarray}
which cannot correctly predict the value of the atmospheric squared mass difference ($\Delta m_{32}^{2}$) in Eq. (\ref{32}). Conversely,  if we use the experimental value of squared mass difference of Eq. (\ref{32}), then Eq. (\ref{mtb}) predicts the absolute values of neutrino mass as follow:
\begin{eqnarray}
\left|m_{1}\right|=0.0247810607~\rm{eV},\nonumber\\
\left|m_{2}\right|=0.0335840950~\rm{eV},\nonumber\\
\left|m_{3}\right|=0.0598990103~\rm{eV},
\end{eqnarray}
which cannot correctly predict the value of $\Delta m_{21}^{2}$ in Eq. (\ref{21}).

\subsection{Neutrino mass matrix from modified BM}
By using Eqs. (\ref{Mo2}), (\ref{Mb}), and (\ref{A}), we have the neutrino mass matrix from the modified bimaximal neutrino mixing matrix as follows:
\begin{eqnarray}
M_{\nu}=\bordermatrix{& & &\cr
&(M_{\nu})_{11} &(M_{\nu})_{12} &(M_{\nu})_{13} \cr
&(M_{\nu})_{21} &(M_{\nu})_{22} &(M_{\nu})_{23} \cr
&(M_{\nu})_{31} &(M_{\nu})_{32} &(M_{\nu})_{33}\cr},
\end{eqnarray}
where
\begin{eqnarray}
(M_{\nu})_{11}=m_{1}\left(\frac{\sqrt{2}c_{x}}{2}-\frac{\sqrt{2}c_{y}s_{x}}{2}\right)^{2}+m_{2}\left(\frac{\sqrt{2}s_{x}}{2}+\frac{\sqrt{2}c_{y}c_{x}}{2}\right)^{2}+m_{3}\frac{s_{y}^{2}}{2},
\end{eqnarray}
\begin{eqnarray}
(M_{\nu})_{12}=m_{1}\left(\frac{\sqrt{2}c_{x}}{2}-\frac{\sqrt{2}c_{y}s_{x}}{2}\right)\left(-\frac{c_{x}}{2}-\frac{c_{y}s_{x}}{2}+\frac{\sqrt{2}s_{y}s_{x}}{2}\right)\nonumber\\+m_{2}\left(\frac{\sqrt{2}s_{x}}{2}+\frac{\sqrt{2}c_{y}c_{x}}{2}\right)\left(-\frac{s_{x}}{2}+\frac{c_{y}c_{x}}{2}-\frac{\sqrt{2}s_{y}c_{x}}{2}\right)\nonumber\\+m_{3}\frac{\sqrt{2}}{2}\left(\frac{s_{y}^{2}}{2}+\frac{\sqrt{2}s_{y}c_{y}}{2}\right),
\end{eqnarray}
\begin{eqnarray}
(M_{\nu})_{13}=m_{1}\left(\frac{\sqrt{2}c_{x}}{2}-\frac{\sqrt{2}c_{y}s_{x}}{2}\right)\left(-\frac{c_{x}}{2}+\frac{c_{y}s_{x}}{2}+\frac{\sqrt{2}s_{y}s_{x}}{2}\right)\nonumber\\+m_{2}\left(\frac{\sqrt{2}s_{x}}{2}+\frac{\sqrt{2}c_{y}c_{x}}{2}\right)\left(\frac{s_{x}}{2}-\frac{c_{y}c_{x}}{2}-\frac{\sqrt{2}s_{y}c_{x}}{2}\right)\nonumber\\+m_{3}\frac{\sqrt{2}}{2}\left(-\frac{s_{y}^{2}}{2}+\frac{\sqrt{2}s_{y}c_{y}}{2}\right),\\
(M_{\nu})_{21}=m_{1}\left(\frac{\sqrt{2}c_{x}}{2}-\frac{\sqrt{2}c_{y}s_{x}}{2}\right)\left(-\frac{c_{x}}{2}-\frac{c_{y}s_{x}}{2}+\frac{\sqrt{2}s_{y}s_{x}}{2}\right)\nonumber\\+m_{2}\left(\frac{\sqrt{2}s_{x}}{2}+\frac{\sqrt{2}c_{y}c_{x}}{2}\right)\left(-\frac{s_{x}}{2}+\frac{c_{y}c_{x}}{2}-\frac{\sqrt{2}s_{y}c_{x}}{2}\right)\nonumber\\+m_{3}\frac{\sqrt{2}}{2}\left(\frac{s_{y}^{2}}{2}+\frac{\sqrt{2}s_{y}c_{y}}{2}\right),\\
(M_{\nu})_{22}=m_{1}\left(-\frac{c_{x}}{2}-\frac{c_{y}s_{x}}{2}+\frac{\sqrt{2}s_{y}s_{x}}{2}\right)^{2}+m_{2}\left(-\frac{s_{x}}{2}+\frac{c_{y}c_{x}}{2}-\frac{\sqrt{2}s_{y}c_{x}}{2}\right)^{2}\nonumber\\+m_{3}\left(\frac{s_{y}}{2}+\frac{\sqrt{2}c_{y}}{2}\right)^{2},\\
(M_{\nu})_{23}=m_{1}\left(-\frac{c_{x}}{2}-\frac{c_{y}s_{x}}{2}+\frac{\sqrt{2}s_{y}s_{x}}{2}\right)\left(\frac{c_{x}}{2}+\frac{c_{y}s_{x}}{2}+\frac{\sqrt{2}s_{y}s_{x}}{2}\right)\nonumber\\+m_{2}\left(\frac{s_{x}}{2}-\frac{c_{y}c_{x}}{2}-\frac{\sqrt{2}s_{y}c_{x}}{2}\right)\left(\frac{s_{x}}{2}-\frac{c_{y}c_{x}}{2}-\frac{\sqrt{2}s_{y}c_{x}}{2}\right)\nonumber\\+m_{3}\left(\frac{s_{y}}{2}+\frac{\sqrt{2}c_{y}}{2}\right)\left(-\frac{s_{y}}{2}+\frac{\sqrt{2}c_{y}}{2}\right),\\
(M_{\nu})_{31}=m_{1}\left(\frac{\sqrt{2}c_{x}}{2}-\frac{\sqrt{2}c_{y}s_{x}}{2}\right)\left(\frac{c_{x}}{2}+\frac{c_{y}s_{x}}{2}+\frac{\sqrt{2}s_{y}s_{x}}{2}\right)\nonumber\\+m_{2}\left(\frac{\sqrt{2}s_{x}}{2}+\frac{\sqrt{2}c_{y}c_{x}}{2}\right)\left(\frac{s_{x}}{2}-\frac{c_{y}c_{x}}{2}-\frac{\sqrt{2}s_{y}c_{x}}{2}\right)\nonumber\\+m_{3}\frac{\sqrt{2}}{2}\left(-\frac{s_{y}^{2}}{2}+\frac{\sqrt{2}s_{y}c_{y}}{2}\right),\\
(M_{\nu})_{32}=m_{1}\left(-\frac{c_{x}}{2}-\frac{c_{y}s_{x}}{2}+\frac{\sqrt{2}s_{y}s_{x}}{2}\right)\left(\frac{c_{x}}{2}+\frac{c_{y}s_{x}}{2}+\frac{\sqrt{2}s_{y}s_{x}}{2}\right)\nonumber\\+m_{2}\left(-\frac{s_{x}}{2}+\frac{c_{y}c_{x}}{2}-\frac{\sqrt{2}s_{y}c_{x}}{2}\right)\left(\frac{s_{x}}{2}-\frac{c_{y}c_{x}}{2}-\frac{\sqrt{2}s_{y}c_{x}}{2}\right)\nonumber\\+m_{3}\left(\frac{s_{y}}{2}+\frac{\sqrt{2}c_{y}}{2}\right)\left(-\frac{s_{y}}{2}+\frac{\sqrt{2}c_{y}}{2}\right),
\end{eqnarray}
\begin{eqnarray}
(M_{\nu})_{33}=m_{1}\left(\frac{c_{x}}{2}+\frac{c_{y}s_{x}}{2}+\frac{\sqrt{2}s_{y}s_{x}}{2}\right)^{2}+m_{2}\left(\frac{s_{x}}{2}-\frac{c_{y}c_{x}}{2}-\frac{\sqrt{2}s_{y}c_{x}}{2}\right)^{2}\nonumber\\+m_{3}\left(-\frac{s_{y}}{2}+\frac{\sqrt{2}c_{y}}{2}\right)^{2},
\end{eqnarray}

For the obtained neutrino mass matrix from modified bimaximal mixing, when we impose the four patterns of two zeros texture in Eqs. (\ref{2233})-(\ref{1223}) and insert the value of $x$ and $y$  in Eq. (\ref{xbm}), then we have:
\begin{eqnarray}
m_{2}=-3473.465412~m_{1},~m_{3}=4.218376~ m_{1},~{\rm{for}}~(M_{\nu})_{22}=(M_{\nu})_{33}=0, \label{mbm}\\
m_{1}=-47615.39155~m_{2},~m_{3}=-655.03754~ m_{2},~{\rm{for}}~(M_{\nu})_{11}=(M_{\nu})_{13}=0, \label{mbm1}\\
m_{1}=0.99999996 ~m_{3},~m_{2}= m_{3},~{\rm{for}}~(M_{\nu})_{12}=(M_{\nu})_{13}=0, \label{mbm2}\\
m_{1}=m_{3},~m_{2}=m_{3},~{\rm{for}}~(M_{\nu})_{12}=(M_{\nu})_{23}=0. \label{mbm3}
\end{eqnarray}
From Eqs. (\ref{mbm})-(\ref{mbm3}), one can see that from four patterns of two zeros texture, for the obtained neutrino mass matrix from modified bimaximal mixing, none of them can give the neutrino mass spectrum which is compatible with the known neutrino mass spectrum.

\subsection{Neutrino mass matrix from modified DC}
By using Eqs. (\ref{Mo}), (\ref{Mb}), and (\ref{A}), we have the neutrino mass matrix from the modified democratic neutrino mixing matrix as follows:
\begin{eqnarray}
M_{\nu}=\bordermatrix{& & &\cr
&(M_{\nu})_{11} &(M_{\nu})_{12} &(M_{\nu})_{13} \cr
&(M_{\nu})_{21} &(M_{\nu})_{22} &(M_{\nu})_{23} \cr
&(M_{\nu})_{31} &(M_{\nu})_{32} &(M_{\nu})_{33}\cr},
\end{eqnarray}
where
\begin{eqnarray}
(M_{\nu})_{11}=m_{1}\left(\frac{\sqrt{2}c_{x}}{2}-\frac{\sqrt{2}c_{y}s_{x}}{2}\right)^{2}+m_{2}\left(\frac{\sqrt{2}s_{x}}{2}+\frac{\sqrt{2}c_{y}c_{x}}{2}\right)^{2}+m_{3}\frac{s_{y}^{2}}{2},
\end{eqnarray}
\begin{eqnarray}
(M_{\nu})_{12}=m_{1}\left(\frac{\sqrt{2}c_{x}}{2}-\frac{\sqrt{2}c_{y}s_{x}}{2}\right)\left(-\frac{\sqrt{6}c_{x}}{6}+\frac{\sqrt{6}c_{y}s_{x}}{6}+\frac{\sqrt{6}s_{y}s_{x}}{3}\right)\nonumber\\+m_{2}\left(\frac{\sqrt{2}s_{x}}{2}+\frac{\sqrt{2}c_{y}c_{x}}{2}\right)\left(-\frac{\sqrt{6}s_{x}}{6}-\frac{\sqrt{6}c_{y}c_{x}}{6}-\frac{\sqrt{6}s_{y}c_{x}}{3}\right)\nonumber\\+m_{3}\frac{\sqrt{2}}{2}\left(-\frac{\sqrt{6}s_{y}^{2}}{6}+\frac{\sqrt{6}s_{y}c_{y}}{3}\right),
\end{eqnarray}
\begin{eqnarray}
(M_{\nu})_{13}=m_{1}\left(\frac{\sqrt{2}c_{x}}{2}-\frac{\sqrt{2}c_{y}s_{x}}{2}\right)\left(-\frac{\sqrt{3}c_{x}}{3}+\frac{\sqrt{3}c_{y}s_{x}}{3}-\frac{\sqrt{3}s_{y}s_{x}}{3}\right)\nonumber\\+m_{2}\left(\frac{\sqrt{2}s_{x}}{2}+\frac{\sqrt{2}c_{y}c_{x}}{2}\right)\left(-\frac{\sqrt{3}s_{x}}{3}+\frac{\sqrt{3}c_{y}c_{x}}{3}+\frac{\sqrt{3}s_{y}c_{x}}{3}\right)\nonumber\\+m_{3}\frac{\sqrt{2}}{2}\left(\frac{\sqrt{3}s_{y}^{2}}{3}-\frac{\sqrt{3}s_{y}c_{y}}{3}\right),
\end{eqnarray}
\begin{eqnarray}
(M_{\nu})_{21}=m_{1}\left(\frac{\sqrt{2}c_{x}}{2}-\frac{\sqrt{2}c_{y}s_{x}}{2}\right)\left(-\frac{\sqrt{6}c_{x}}{6}+\frac{\sqrt{6}c_{y}s_{x}}{6}+\frac{\sqrt{6}s_{y}s_{x}}{3}\right)\nonumber\\+m_{2}\left(\frac{\sqrt{2}s_{x}}{2}+\frac{\sqrt{2}c_{y}c_{x}}{2}\right)\left(-\frac{\sqrt{6}s_{x}}{6}-\frac{\sqrt{6}c_{y}c_{x}}{6}-\frac{\sqrt{6}s_{y}c_{x}}{3}\right)\nonumber\\+m_{3}\frac{\sqrt{2}}{2}\left(-\frac{\sqrt{6}s_{y}^{2}}{6}+\frac{\sqrt{6}s_{y}c_{y}}{3}\right),
\end{eqnarray}
\begin{eqnarray}
(M_{\nu})_{22}=m_{1}\left(-\frac{\sqrt{6}c_{x}}{6}+\frac{\sqrt{6}c_{y}s_{x}}{6}+\frac{\sqrt{6}s_{y}s_{x}}{3}\right)^{2}\nonumber\\+m_{2}\left(-\frac{\sqrt{6}s_{x}}{6}-\frac{\sqrt{6}c_{y}c_{x}}{6}-\frac{\sqrt{6}s_{y}c_{x}}{3}\right)^{2}\nonumber\\+m_{3}\left(-\frac{\sqrt{6}s_{y}}{6}+\frac{\sqrt{6}c_{y}}{3}\right)^{2},
\end{eqnarray}
\begin{eqnarray}
(M_{\nu})_{23}=m_{1}\left(\frac{\sqrt{6}c_{x}}{6}-\frac{\sqrt{6}c_{y}s_{x}}{6}-\frac{\sqrt{6}s_{y}s_{x}}{3}\right)\left(\frac{\sqrt{3}c_{x}}{3}+\frac{\sqrt{3}c_{y}s_{x}}{3}+\frac{\sqrt{3}s_{y}s_{x}}{3}\right)\nonumber\\+m_{2}\left(-\frac{\sqrt{6}s_{x}}{6}-\frac{\sqrt{6}c_{y}c_{x}}{6}-\frac{\sqrt{6}s_{y}c_{x}}{3}\right)\left(-\frac{\sqrt{3}s_{x}}{3}+\frac{\sqrt{3}c_{y}c_{x}}{3}+\frac{\sqrt{3}s_{y}c_{x}}{3}\right)\nonumber\\+m_{3}\left(-\frac{\sqrt{6}s_{y}}{6}+\frac{\sqrt{6}c_{y}}{3}\right)\left(\frac{\sqrt{3}s_{y}}{3}-\frac{\sqrt{3}c_{y}}{3}\right),
\end{eqnarray}
\begin{eqnarray}
(M_{\nu})_{31}=m_{1}\left(\frac{\sqrt{2}c_{x}}{2}-\frac{\sqrt{2}c_{y}s_{x}}{2}\right)\left(-\frac{\sqrt{3}c_{x}}{3}-\frac{\sqrt{3}c_{y}s_{x}}{3}-\frac{\sqrt{3}s_{y}s_{x}}{3}\right)\nonumber\\+m_{2}\left(\frac{\sqrt{2}s_{x}}{2}+\frac{\sqrt{2}c_{y}c_{x}}{2}\right)\left(-\frac{\sqrt{3}s_{x}}{3}+\frac{\sqrt{3}c_{y}c_{x}}{3}+\frac{\sqrt{3}s_{y}c_{x}}{3}\right)\nonumber\\+m_{3}\frac{\sqrt{2}}{2}\left(\frac{\sqrt{3}s_{y}^{2}}{3}-\frac{\sqrt{3}s_{y}c_{y}}{3}\right),
\end{eqnarray}
\begin{eqnarray}
(M_{\nu})_{32}=m_{1}\left(-\frac{\sqrt{6}c_{x}}{6}+\frac{\sqrt{6}c_{y}s_{x}}{6}+\frac{\sqrt{6}s_{y}s_{x}}{3}\right)\left(-\frac{\sqrt{3}c_{x}}{3}-\frac{\sqrt{3}c_{y}s_{x}}{3}-\frac{\sqrt{3}s_{y}s_{x}}{3}\right)\nonumber\\+m_{2}\left(-\frac{\sqrt{6}s_{x}}{6}-\frac{\sqrt{6}c_{y}c_{x}}{6}-\frac{\sqrt{6}s_{y}c_{x}}{3}\right)\left(-\frac{\sqrt{3}s_{x}}{3}+\frac{\sqrt{3}c_{y}c_{x}}{3}+\frac{\sqrt{3}s_{y}c_{x}}{3}\right)\nonumber\\+m_{3}\left(-\frac{\sqrt{6}s_{y}}{6}+\frac{\sqrt{6}c_{y}}{3}\right)\left(\frac{\sqrt{3}s_{y}}{3}-\frac{\sqrt{3}c_{y}}{3}\right),
\end{eqnarray}
\begin{eqnarray}
(M_{\nu})_{33}=m_{1}\left(-\frac{\sqrt{3}c_{x}}{3}-\frac{\sqrt{3}c_{y}s_{x}}{3}+\frac{\sqrt{3}s_{y}s_{x}}{3}\right)^{2}\nonumber\\+m_{2}\left(-\frac{\sqrt{3}s_{x}}{3}+\frac{\sqrt{3}c_{y}c_{x}}{3}+\frac{\sqrt{3}s_{y}c_{x}}{3}\right)^{2}\nonumber\\+m_{3}\left(-\frac{\sqrt{3}s_{y}}{3}-\frac{\sqrt{3}c_{y}}{3}\right)^{2},
\end{eqnarray}

For the obtained neutrino mass matrix from the modified democratic mixing, if we impose the four patterns of two zeros texture in Eqs. (\ref{2233})-(\ref{1223}) and insert the value of $x$ and $y$ in Eq. (\ref{xdc}), then we have:
\begin{eqnarray}
m_{1}=-1.41771494~m_{3},~m_{2}=-0.66976966~ m_{3},~{\rm{for}}~(M_{\nu})_{22}=(M_{\nu})_{33}=0, \label{mdc}\\
m_{2}=-3.275173358~m_{1},~m_{3}=8.727515495~ m_{1},~{\rm{for}}~(M_{\nu})_{11}=(M_{\nu})_{13}=0, \label{mdc1}\\
m_{1}=0.999999999 ~m_{3},~m_{2}=0.999999999~m_{3},~{\rm{for}}~(M_{\nu})_{12}=(M_{\nu})_{13}=0, \label{mdc2}\\
m_{1}=m_{2},~m_{3}=0.999999999~m_{2},~{\rm{for}}~(M_{\nu})_{12}=(M_{\nu})_{23}=0. \label{mdc3}
\end{eqnarray}
From Eqs. (\ref{mdc})-(\ref{mdc3}), one can see that from four patterns of two zeros texture for the obtained neutrino mass matrix from the modified democratic mixing, only the neutrino mass matrix with two zeros texture: $(M_{\nu})_{11}=(M_{\nu})_{13}=0$ can predicts the correct neutrino mass spectrum:
\begin{eqnarray}
\left|\frac{m_{2}}{m_{1}}\right|,~\left|\frac{m_{3}}{m_{2}}\right|>1,
 \label{IH}
\end{eqnarray}
which imply that the neutrino mass is normal hierarchy: $\left|m_{1}\right|<\left|m_{2}\right|<\left|m_{3}\right|$.  

By using the experimental values of squared mass difference as shown in Eqs. (\ref{21}), we obtain the absolute values of neutrino mass for modified democratic neutrino mixing matrix as follow:
\begin{eqnarray}
\left|m_{1}\right|=0.0027934235~\rm{eV},\nonumber\\
\left|m_{2}\right|=0.0091489461~\rm{eV},\nonumber\\
\left|m_{3}\right|=0.0244075808~\rm{eV}.\label{dcm}
\end{eqnarray}
The neutrino masses in Eq. (\ref{dcm}) cannot correctly predict the squared mass difference for atmospheric neutrino $\Delta m_{32}^{2}$ of Eq. (\ref{32}).  Conversely, if we first use $\Delta m_{32}^{2}$ in Eq. (\ref{32}) for determining $m_{1}, m_{2}$, and $m_{3}$, then we have:
\begin{eqnarray}
\left|m_{1}\right|=0.0061229117~\rm{eV},\nonumber\\
\left|m_{2}\right|=0.0200535970~\rm{eV},\nonumber\\
\left|m_{3}\right|=0.0534990351~\rm{eV},\label{dcm1}
\end{eqnarray}
which cannot predict corectly the squared mass difference for solar neutrino $\Delta m_{21}^{2}$ in Eq. (\ref{21}).

\section{Conclusion}
The modified neutrino mixing matrices (TB, BM, DC) are obtained by introducing a perturbation matrices into neutrino mixing matrices.  All of the modified neutrino mixing matrices can give nonzero $\theta_{13}$.  Even though all of the modified neutrino mixing matrices can give nonzero $\theta_{13}$, but only the modified DC neutrino mixing  matrix can predict the value of $\theta_{13}$ which is compatible with the latest experimental results. The modified TB neutrino mixing matrix predicts the value of $\theta_{13}$ greater than the upper bound value of T2K experiment, meanwhile the modified BM neutrino mixing matrix predicts the value of $\theta_{13}$ below the lower bound of T2K experiment.  When the two zeros texture to be imposed on the obtained neutrino mass matrices from modified mixing matrices, only the obtained neutrino mass matrices from modified TB with two zeros texture $(M_{\nu})_{22}=(M_{\nu})_{33}=0$ and the modified DC with two zeros texture $(M_{\nu})_{11}=(M_{\nu})_{13}=0$ can give the neutrino mass spectrum in agreement with one of the known neutrino mass spectrum, that is normal hierarchy: $\left|m_{1}\right|< \left|m_{2}\right|<\left|m_{3}\right|$ .  If we use the experimental results of squared mass difference $\Delta m_{21}^{2}$ to obtain the values of neutrino masses, then the obtained neutrino masses cannot predict the correct value of $\Delta m_{32}^{2}$.   Conversely, if we use the experimental value of squared mass difference $\Delta m_{32}^{2}$ to obtain the values of neutrino masses, then the obtained neutrino masses cannot correctly predict the correct $\Delta m_{21}^{2}$. 

\section*{Acknowledgment}
Author thank to  reviewer(s), the final version of this manuscript is the result of the first manuscript that has changed substantially due to the reviewer(s) comments and suggestions.

\end{document}